\documentclass[11pt]{article}
\usepackage{amsmath,amssymb,bm}
\usepackage{graphicx}
\usepackage[margin=2.5cm]{geometry}
\usepackage[colorlinks=true,linkcolor=blue,citecolor=blue,urlcolor=blue]{hyperref}
\usepackage{microtype}

\newcommand{\KLC}{Kerr--Levi-Civita}

\newcommand{\dd}{\mathrm{d}}
\newcommand{\Dr}{\Delta_r}
\newcommand{\Dx}{\Delta_x}
\newcommand{\Ecal}{\mathcal{E}}
\newcommand{\Ccal}{\mathcal{C}}
\begin{document}

\title{Charged Kerr--Levi-Civita geometries in Einstein--Maxwell and\\
low-energy heterotic string theory}
\author{Haryanto M. Siahaan\thanks{haryanto.siahaan@unpar.ac.id}\\[4pt]
\small Program Studi Fisika, Universitas Katolik Parahyangan,\\
\small Bandung 40141, Indonesia}
\date{}
\maketitle

\begin{abstract}
We construct and compare two charged rotating extensions of the Kerr--Levi-Civita geometry. In Einstein--Maxwell theory, a fixed magnetic Kerr--Newman Ernst representative is inverted, and covariance of the coupled Ernst equations, integrability of the dragging and electric-potential quadratures, and the field equations are verified. The resulting local line element agrees with a strong-field representative obtained independently, but no global equivalence of the azimuthal quotients is assumed. In low-energy heterotic string theory, the Hassan--Sen map is instead applied after the vacuum Kerr--Levi-Civita inversion, generating Maxwell, dilaton, and Kalb--Ramond fields. Both branches possess regular local Killing horizons. For the Einstein--Maxwell branch, an exact denominator factorization proves that the subextreme exterior contains neither Ernst zeros nor azimuthal closed timelike curves. The former Kerr ring has finite curvature, although it lies inside an interior region of timelike azimuthal orbits. Its Kretschmann scalar has the form $8\mathcal P_{\rm N}/\mathcal H^6$ and approaches a parameter-independent Levi-Civita law at fixed off-axis latitude. The heterotic branch is qualitatively different: the real component connected to the horizon terminates at a finite-radius $\Lambda=0$ surface. Exact string-frame slice factorizations and independent Einstein-frame calculations show that this surface is a curvature singularity rather than a conformal-frame artifact. Both families are generically Petrov type I. The results distinguish exact local solution generation from the unresolved construction of complete global spacetimes and the identification of possible distributional sources.
\end{abstract}


\section{Introduction}
\label{sec:intro}
Stationary and axisymmetric exact solutions remain one of the clearest
laboratories for nonlinear general relativity.  The Kerr and Kerr--Newman
geometries \cite{Kerr1963,Newman1965} are the canonical rotating vacuum and
electrovacuum examples, while the Ernst formulation converts the reduced
field equations into a complex-potential system
\cite{Ernst1968a,Ernst1968b}.  Ehlers, Harrison, Geroch, and Kinnersley
transformations then organize large families of new solutions from simple
seeds \cite{Ehlers1962,Harrison1968,Geroch1971,Kinnersley1973}; a broad
account of this exact-solution framework is given in
Ref.~\cite{Stephani2003}.
The Levi-Civita (LC) metric is the canonical static cylindrically symmetric
vacuum solution \cite{LeviCivita1919}.  Its non-flat asymptotic structure
makes it a model of a gravitating environment rather than an isolated source.
Reciprocal transformations, beginning with Buchdahl's construction
\cite{Buchdahl1954} and its modern rotating and matter extensions
\cite{Barrientos2024Buchdahl}, have recently placed compact objects in this
background.  The Schwarzschild--Levi-Civita geometry was studied in
Refs.~\cite{Mazharimousavi2025PLB,Amirabi2025}; the rotating
\KLC{} (KLC) spacetime was generated from Kerr by a discrete inversion of the
magnetic Ernst potential \cite{Barrientos2025KerrLC}; and the same route has
been extended to a NUT-charged seed with a Manko--Ruiz frame parameter
\cite{Siahaan:2026knlc}.
The wider context is the rapidly developing theory of black holes in
non-asymptotically-flat external backgrounds.  The Melvin universe and
Ernst's magnetized black holes \cite{Melvin1964,Ernst1976} already show that
rotation, charge, and the choice of asymptotic Killing frame can radically
alter ergoregions \cite{Gibbons2013}.  More recent Ehlers--Harrison
constructions include swirling universes \cite{Astorino2022Swirling}, mixed
electromagnetic-swirling type-I backgrounds
\cite{Barrientos2024Swirling}, and Kerr--Newman black holes in a combined
Melvin-swirling environment \cite{DiPinto2025}.  Harrison-generated
Bertotti--Robinson--Bonnor--Melvin backgrounds have also been analysed
thermodynamically \cite{Astorino2025BRBM}.  Astorino's general classification
then recovers the same local charged KLC line element within a strong-field
sector of this broader background family \cite{Astorino:2026backgrounds}.
The inherited azimuthal identification must be compared separately before a
global equivalence of quotient spacetimes can be asserted.
The vacuum KLC geometry is remarkable because inversion removes the
polynomial-curvature singularity associated with the Kerr ring while
preserving the Kerr horizon polynomial.  This raises two questions.  First,
does this regularization survive when the seed carries a genuine matter
stress tensor?  Second, is the answer specific to Einstein--Maxwell theory,
or does it persist when electric charge is accompanied by a dilaton and an
antisymmetric tensor field?
There are two natural and technically different ways to address these
questions.  In Einstein--Maxwell theory the magnetic
Weyl--Lewis--Papapetrou (WLP) reduction is governed by the coupled Ernst pair
$(\Ecal,\Phi)$.  The vacuum inversion extends to
\begin{equation}
 \Ecal\longmapsto\frac{1}{\Ecal},
 \qquad
 \Phi\longmapsto\frac{\Phi}{\Ecal},
 \label{eq:inversion-intro}
\end{equation}
so one may fix the standard Kerr--Newman Ernst representative and immerse it.
This operation must not be conflated with applying the bare Harrison map to
Kerr and then inverting: inversion acts on Ernst potentials rather than on
electromagnetic gauge-equivalence classes, and it conjugates a constant
potential shift into a Harrison transformation.  The resulting ordering and
seed-gauge subtlety are established exactly in
Section~\ref{sec:gauge-harrison-order}.  Inversion and Harrison operations have
recently been used in other vacuum/electrovacuum constructions as well
\cite{Barrientos:2026inversion}.  Astorino has already shown that a
Kerr--Newman--Levi-Civita metric is contained in the large-field limit of
Kerr--Newman--Melvin and has explained how inversion arises from standard
Ehlers/Harrison operations \cite{Astorino:2026backgrounds}.  We therefore do
\emph{not} claim the first existence of the Einstein--Maxwell metric.  The
contribution of the present route is instead a self-contained direct Ernst
construction: a proof of \eqref{eq:inversion-intro} for the coupled equations,
compact rational potentials, explicit integrability of both quadratures, a
full field-equation check, an explicit static member, and several geometric
results not apparent from the strong-field limiting representation.
The heterotic branch belongs to the lineage of charged dilatonic black holes
initiated by the Gibbons--Maeda and Garfinkle--Horowitz--Strominger solutions
\cite{GibbonsMaeda1988,GHS1991}.  Its antisymmetric tensor is the
Kalb--Ramond field \cite{KalbRamond1974}.  The appropriate stationary charging
map is the Hassan--Sen transformation \cite{HassanSen1992,Sen1992}; broader
Ehlers--Harrison-type symmetry groups are also known in
Einstein--Maxwell--dilaton--axion gravity \cite{Galtsov1994}.  Hassan--Sen acts
on a \emph{vacuum} stationary seed, so the order is reversed: one first
constructs KLC and then charges it.  Rotation activates the complete
four-dimensional field content.  In addition to the metric and Maxwell
potential, the image contains a nonconstant dilaton and a Kalb--Ramond
two-form whose three-form field strength includes a gauge Chern--Simons term.
The static truncation was recently used to obtain a charged
Schwarzschild--Levi-Civita solution \cite{Mazharimousavi:2026xbp}; the rotating
case is qualitatively richer because the antisymmetric tensor no longer
vanishes.
The main findings force a distinction between \emph{local exact solutions}
and \emph{global black-hole spacetimes}.  The Einstein--Maxwell branch has a
regular outer Killing horizon and a regular former Kerr ring, but charge
creates an interior region with $g_{\phi\phi}<0$.  Its exterior nevertheless
has neither Ernst zeros nor azimuthal closed timelike curves.  The heterotic
branch also has a regular local Killing horizon, but its Hassan--Sen factor
changes sign at finite radius in every fixed off-axis direction.  The real
dilaton branch therefore terminates at a singular $\Lambda=0$ wall before a
full LC asymptotic end is reached.  For this reason we use the neutral word
``geometries'' in the title and reserve ``black hole'' for local horizon
statements or for a global completion once such a completion has been
established.
The distinction is also important because a componentwise proof of the field
equations does not determine the maximal extension or exclude distributional
sources.  Herdeiro and Novo recently showed, for two different static
symmetry-generated electrovacuum geometries, that a chart which is pointwise
vacuum can conceal an annular source visible in Weyl coordinates
\cite{HerdeiroNovo2026}.  Their result does not establish such a source for
the present rotating families, but it demonstrates why the global Weyl/rod
analysis must be kept separate from the local computer-algebra verification.
The paper is organized as follows.  Section~\ref{sec:seed} fixes the magnetic
WLP conventions and reviews the KLC seed.  Section~\ref{sec:KNLC} proves the
electrovacuum inversion, clarifies its Harrison/gauge conjugacy and ordering,
and constructs the Einstein--Maxwell branch; Section~\ref{sec:KSLC} constructs
the heterotic branch.  The static electrovacuum member is isolated in
Section~\ref{sec:staticKNLC}.  Section~\ref{sec:horizons} gives the local
horizon mechanics and the global azimuthal normalization.
Section~\ref{sec:singularities} analyses the Ernst denominator, the former
ring, the heterotic singular wall, and the Petrov type.
Section~\ref{sec:scope} compares the branches and states precisely the scope
of novelty and of the global claims.  Technical proofs and explicit
polynomials are collected in the appendices.  We use signature $(-,+,+,+)$ and coordinates $(t,r,x,\phi)$ with
$x=\cos\theta$-type.  All dimensional quantities are measured relative to the
fixed inversion scale $\ell$ introduced in Sec.~\ref{sec:inversion-scale}.
Because the geometries are not asymptotically flat, $m,a,q,b$, and the
Hassan--Sen parameter $s$ are generating parameters unless a separate charge
prescription is specified.
\section{The \KLC{} seed and conventions}
\label{sec:seed}
\subsection{Normalization scale of the inversion}
\label{sec:inversion-scale}
Throughout, all coordinates and parameters have been rendered dimensionless
with a fixed reference length $\ell$ ($r_{\rm phys}=\ell r$,
$m_{\rm phys}=\ell m$, and so on, with
$\dd s^{2}_{\rm phys}=\ell^{2}\dd s^{2}$), and the inversion is applied to
the correspondingly reduced dimensionless Ernst potentials.  Physical areas,
Maxwell flux charges, and curvature scalars follow from the displayed ones
upon multiplication by $\ell^{2}$, $\ell$, and $\ell^{-4}$, respectively;
this is what renders quantities such as the conicity factors pure numbers.
The convention plays no role in the verification of the field equations.
We use the magnetic WLP form adapted to the axial Killing vector
$\partial_\phi$,
\begin{equation}
 \dd s^{2}=f(\dd\phi-\omega\,\dd t)^{2}
 -\frac{\varrho^{2}}{f}\,\dd t^{2}
 +\frac{e^{2\gamma}}{f}
 \left(\frac{\dd r^{2}}{\Dr}+\frac{\dd x^{2}}{\Dx}\right),
 \qquad
 \varrho^{2}=\Dr\Dx,
 \qquad
 \Dx=1-x^{2}.
 \label{eq:wlp}
\end{equation}
Thus $f=g_{\phi\phi}$ and $\omega=-g_{t\phi}/g_{\phi\phi}$.  In vacuum the
Ernst potential is written $\Ecal=f-i\chi$, with twist equations
\begin{equation}
 \partial_x\chi=\frac{f^{2}}{\Dx}\partial_r\omega,
 \qquad
 \partial_r\chi=-\frac{f^{2}}{\Dr}\partial_x\omega.
 \label{eq:vac-twist}
\end{equation}
For the Kerr seed define
\begin{equation}
 \Dr=r^{2}-2mr+a^{2},\qquad
 \Sigma=r^{2}+a^{2}x^{2},\qquad
 Q=r^{2}+a^{2},\qquad
 N=Q^{2}-a^{2}\Dr\Dx.
 \label{eq:vac-blocks}
\end{equation}
Its magnetic WLP data are
\begin{equation}
 \begin{aligned}
 f_0&=\frac{\Dx N}{\Sigma},
 &\qquad \omega_0&=\frac{2mar}{N},\\
 e^{2\gamma}&=f_0\Sigma=\Dx N,
 &\qquad \chi_0&=\frac{2amx\big[(x^{2}-3)r^{2}-a^{2}(1+x^{2})\big]}{\Sigma}.
 \end{aligned}
 \label{eq:seeddata}
\end{equation}
The KLC image is obtained from $\Ecal_0=f_0-i\chi_0$ by
$\Ecal_0\mapsto\Ecal_0^{-1}$.  Set
\begin{equation}
 W\equiv|\Ecal_0|^{2}=f_0^{2}+\chi_0^{2}.
 \label{eq:Wvac}
\end{equation}
Then $f_N=f_0/W$, $\chi_N=-\chi_0/W$, while $e^{2\gamma}$ and
$\varrho^{2}$ remain invariant.  The inverse twist equations determine a
rational function $\omega_N$.  The metric components are
\begin{equation}
\begin{aligned}
 \widetilde g_{\phi\phi}&=f_N,
 &\widetilde g_{t\phi}&=-f_N\omega_N,
 &\widetilde g_{tt}&=f_N\omega_N^{2}-\frac{\Dr\Dx}{f_N},\\
 \widetilde g_{rr}&=\frac{W\Sigma}{\Dr},
 &\widetilde g_{xx}&=\frac{W\Sigma}{\Dx}.
\end{aligned}
 \label{eq:KLCcomponents}
\end{equation}
Two identities used repeatedly below are
\begin{equation}
 \widetilde g_{tt}\widetilde g_{\phi\phi}
 -\widetilde g_{t\phi}^{2}=-\Dr\Dx,
 \qquad
 \widetilde g_{rr}\Dr=\widetilde g_{xx}\Dx=W\Sigma.
 \label{eq:KLCidentities}
\end{equation}
In particular, the local Killing horizons are at
$r_\pm=m\pm\sqrt{m^{2}-a^{2}}$.
The azimuthal normalization is global data and must not be hidden in local
formulae.  On the surfaces of fixed $t$ and $r$, let
$R=\int\sqrt{\widetilde g_{xx}}\,\dd x$ denote the proper distance from the
axis; from \eqref{eq:KLCcomponents},
$R^{2}=16a^{2}m^{2}(r^{2}+a^{2})\,\Dx\,[1+O(\Dx)]$ near $x=\pm1$, because
the axis value of the seed twist is $\chi_0|_{x=\pm1}=\mp4am$.  In the
algebraic coordinate $\phi$, the transverse metric near either axis then
reads
\begin{equation}
 \dd s_{\perp}^{2}=\dd R^{2}+\frac{R^{2}}{\Ccal_0^{2}}\dd\phi^{2}
 +O(R^{4}),
 \qquad
 \Ccal_0=16a^{2}m^{2}.
 \label{eq:KLCconicity}
\end{equation}
The coefficient $\Ccal_0$ is independent of $r$, so the defect is uniform
along the axis and a single global period choice suffices.  Hence one may
either keep an arbitrary period $\Delta\phi$ and retain the
corresponding conical defect, or define $\phi=\Ccal_0\varphi$ with
$\varphi\sim\varphi+2\pi$.  Equivalently, the regular period of the original
coordinate is $\Delta\phi_{\rm reg}=2\pi\Ccal_0$
\cite{Barrientos2025KerrLC}.  We keep $\Delta\phi$ explicit until
Section~\ref{sec:horizons}.
\section{Einstein--Maxwell sector}
\label{sec:KNLC}
\subsection{Electrovacuum inversion}
\label{sec:inversion}
Let
\begin{equation}
 f=\operatorname{Re}\Ecal+\varepsilon|\Phi|^{2},
 \qquad
 \bm J=\nabla\Ecal+2\varepsilon\bar\Phi\nabla\Phi,
 \qquad
 \varepsilon=\pm1.
 \label{eq:ernstdefs}
\end{equation}
The coupled Ernst equations are
\begin{equation}
 f\nabla^{2}\Ecal=\bm J\mathbin{\cdot}\nabla\Ecal,
 \qquad
 f\nabla^{2}\Phi=\bm J\mathbin{\cdot}\nabla\Phi,
 \label{eq:ernsteqs}
\end{equation}
where the dot is the complex-bilinear, not Hermitian, product on the flat Weyl
base.  Our magnetic convention is $\varepsilon=-1$, so
$f=\operatorname{Re}\Ecal-|\Phi|^{2}$.
\begin{quote}
\textbf{Proposition 1.}
If $(\Ecal_0,\Phi_0)$ solves \eqref{eq:ernsteqs}, then
\begin{equation}
 \Ecal=\frac{1}{\Ecal_0},
 \qquad
 \Phi=\frac{e^{i\alpha}\Phi_0}{\Ecal_0},
 \qquad \alpha\in\mathbb R,
 \label{eq:inversion-full}
\end{equation}
solves the same equations for either value of $\varepsilon$, and
$f=f_0/|\Ecal_0|^{2}$.
\end{quote}
The proof is given in Appendix~\ref{app:proof}.  The phase is the usual
Maxwell duality rotation; replacing it by a constant of non-unit modulus does
not preserve the equations.  The inversion is a projective $SU(2,1)$ action
and can also be represented as a composition, or limiting case, of standard
Ehlers/Harrison transformations
\cite{Ehlers1962,Harrison1968,Kinnersley1973,Astorino:2026backgrounds}.
\subsection{Gauge shifts, Harrison conjugacy, and ordering}
\label{sec:gauge-harrison-order}
Before fixing the Kerr--Newman seed, one must specify which representative of
its electromagnetic potentials is to be inverted.  In the magnetic
convention $f=\operatorname{Re}\Ecal-|\Phi|^{2}$, define the constant
potential shift
\begin{equation}
 \mathsf D_c:(\Ecal,\Phi)\longmapsto
 \left(\Ecal+2\bar c\,\Phi+|c|^{2},\,\Phi+c\right),
 \qquad c\in\mathbb C,
 \label{eq:gauge-shift}
\end{equation}
and the Harrison transformation
\begin{equation}
 \mathsf H_c:(\Ecal,\Phi)\longmapsto
 \left(\frac{\Ecal}{\Delta_c},\,
 \frac{\Phi+c\Ecal}{\Delta_c}\right),
 \qquad
 \Delta_c=1+2\bar c\,\Phi+|c|^{2}\Ecal.
 \label{eq:harrison-map}
\end{equation}
On each simply connected chart, the map $\mathsf D_c$ leaves $f$, the
four-metric, and the Maxwell field strength unchanged; its real and imaginary
parts shift the two reduced Maxwell scalar potentials by constants.  For a
periodic azimuthal coordinate, a constant real shift of $A_\phi$ may retain
global Wilson-line information, so ``gauge-equivalent'' in this subsection
refers to the local field configuration.  These maps are elements of the
projective $SU(2,1)$ solution-generating action
\cite{Harrison1968,Kinnersley1973,Astorino:2026backgrounds}.
A direct substitution gives
\begin{align}
 \mathsf I\!\left[\mathsf H_c(\Ecal,\Phi)\right]
 &=\left(\frac{\Delta_c}{\Ecal},\,
 \frac{\Phi+c\Ecal}{\Ecal}\right)
 \nonumber\\
 &=\left(\frac{1}{\Ecal}
 +2\bar c\frac{\Phi}{\Ecal}+|c|^{2},\,
 \frac{\Phi}{\Ecal}+c\right)
 =\mathsf D_c\!\left[\mathsf I(\Ecal,\Phi)\right].
 \label{eq:intertwining-proof}
\end{align}
Since inverting twice returns every potential pair unchanged, this proves the exact intertwining and
conjugacy relations
\begin{equation}
 \mathsf I\circ\mathsf H_c=\mathsf D_c\circ\mathsf I,
 \qquad
 \mathsf H_c=\mathsf I\circ\mathsf D_c\circ\mathsf I,
 \qquad
 \mathsf H_c\circ\mathsf I=\mathsf I\circ\mathsf D_c .
 \label{eq:intertwining}
\end{equation}
Thus inversion converts a locally field-trivial constant shift on one
representative into a generally nontrivial Harrison deformation of the
inverted representative.  In particular, inversion does not descend to the
quotient of Ernst potentials by constant electromagnetic-potential shifts;
the seed representative must be fixed before applying $\mathsf I$.
The two orderings are genuinely noncommuting.  On the real section, suppose
that a parameter redefinition could make
$\mathsf H_{\hat b}\circ\mathsf I=\mathsf I\circ\mathsf H_b$ as maps.  Their
gravitational Ernst components would obey
\begin{equation}
 \frac{1}{\Ecal+2\hat b\Phi+\hat b^{2}}
 =\frac{1+2b\Phi+b^{2}\Ecal}{\Ecal},
 \label{eq:ordering-real}
\end{equation}
so that
\begin{equation}
 \Ecal=(\Ecal+2\hat b\Phi+\hat b^{2})
 (1+2b\Phi+b^{2}\Ecal).
 \label{eq:ordering-polynomial}
\end{equation}
The coefficient of $\Phi^{2}$ is $4b\hat b$.  If either parameter vanishes,
the remaining coefficients force the other to vanish as well.  Hence the two
maps agree identically only in the trivial case $b=\hat b=0$.
For a vacuum seed $(\Ecal_0,0)$, charging after inversion is equivalently the
inversion of a constant-shifted vacuum representative:
\begin{equation}
 \mathsf H_b\circ\mathsf I(\Ecal_0,0)
 =\mathsf I\circ\mathsf D_b(\Ecal_0,0)
 =\left(\frac{1}{\Ecal_0+b^{2}},\,
 \frac{b}{\Ecal_0+b^{2}}\right),
 \qquad b\in\mathbb R.
 \label{eq:harrison-dressed-klc}
\end{equation}
Although $\mathsf D_b(\Ecal_0,0)$ is locally field-equivalent to the
original vacuum seed, its inverted image has a nonconstant electromagnetic
potential
and is a genuine Einstein--Maxwell solution.  Conversely,
\begin{equation}
 \mathsf I\circ\mathsf H_b(\Ecal_0,0)
 =\mathsf D_b\circ\mathsf I(\Ecal_0,0)
 \label{eq:bare-harrison-first}
\end{equation}
is only a constant-shifted vacuum KLC solution.  The KNLC family constructed
below must therefore be understood as the inversion of the fixed standard
Kerr--Newman Ernst pair, not as the inversion of the bare Harrison image of
Kerr in Eq.~\eqref{eq:harrison-map}.  Putting a bare Harrison image into a
standard Kerr--Newman parametrization may require further normalizations,
gauge choices, coordinate rescalings, and parameter redefinitions, none of
which is part of the identity \eqref{eq:intertwining}.
The auxiliary branch \eqref{eq:harrison-dressed-klc} already has a useful
global diagnostic.  For the Kerr seed, Eq.~\eqref{eq:seeddata} gives
$\Ecal_0|_{x=\pm1}=\pm4iam$.  Writing
$\Phi_b=A_\phi^{(b)}+i\widetilde A_\phi^{(b)}$ and defining
\begin{equation}
 \Ccal_b=b^{4}+16a^{2}m^{2},
 \label{eq:Cb}
\end{equation}
one finds
\begin{equation}
 \left.A_\phi^{(b)}\right|_{x=\pm1}=\frac{b^{3}}{\Ccal_b},
 \qquad
 \left.\widetilde A_\phi^{(b)}\right|_{x=\pm1}
 =\mp\frac{4abm}{\Ccal_b}.
 \label{eq:harrison-dressed-axis}
\end{equation}
Because $e^{2\gamma}$ and $\varrho^{2}$ are unchanged, the same axis expansion
as in Eq.~\eqref{eq:generic-conicity} gives
$\Delta\phi_{\rm reg}^{(b)}=2\pi\Ccal_b$.  The corresponding Maxwell fluxes
are
\begin{equation}
 \mathcal Q_{\rm m}^{(b)}=0,
 \qquad
 \mathcal Q_{\rm e}^{(b)}
 =\frac{\Delta\phi}{2\pi}\frac{4abm}{\Ccal_b},
 \qquad
 \left.\mathcal Q_{\rm e}^{(b)}\right|_{\rm reg}=4abm.
 \label{eq:harrison-dressed-charge}
\end{equation}
Thus the Harrison-dressed KLC branch is locally nonvacuum and, for $abm\ne0$,
carries nonzero electric flux, while retaining the vacuum Kerr polynomial
$\Dr=r^{2}-2mr+a^{2}$.  This last feature is structurally shared by the
heterotic branch, but the analogy is only organizational: the Hassan--Sen map
belongs to a different theory and additionally generates the dilaton and
Kalb--Ramond sectors.  We use the branch here to clarify the ordering and
seed-gauge issue; a complete invariant and global comparison with KNLC and
with known magnetized or swirling solutions is left for separate work.
\subsection{Kerr--Newman seed and transformed fields}
For Kerr--Newman, now with
\begin{equation}
 \Dr=r^{2}-2mr+a^{2}+q^{2},
 \qquad
 N=(r^{2}+a^{2})^{2}-a^{2}\Dr\Dx,
 \label{eq:charged-blocks}
\end{equation}
the magnetic Ernst pair is
\begin{equation}
 \Ecal_0=f_0+|\Phi_0|^{2}-i\chi_0,
 \qquad
 \Phi_0=\frac{q(rx-ia)}{ax+ir},
 \label{eq:KNpotentials}
\end{equation}
where $f_0=\Dx N/\Sigma$ and
\begin{equation}
 \chi_0=-\frac{2ax\big[(a^{2}-r^{2})mx^{2}+q^{2}rx^{2}
 +a^{2}m+3mr^{2}-q^{2}r\big]}{\Sigma}.
 \label{eq:KNchi}
\end{equation}
It is useful to introduce
\begin{align}
 \mathcal U&=\Dx N+q^{2}(r^{2}x^{2}+a^{2}),
 \nonumber\\
 \mathcal N&=-2ax\big[(a^{2}-r^{2})mx^{2}+q^{2}rx^{2}
 +a^{2}m+3mr^{2}-q^{2}r\big],
 \label{eq:UNW}\\
 \mathcal W&=\mathcal U^{2}+\mathcal N^{2}
 =\Sigma^{2}|\Ecal_0|^{2}.
 \nonumber
\end{align}
The inverted potentials are
\begin{equation}
 \begin{aligned}
 f_N&=\frac{\Dx N\Sigma}{\mathcal W},
 &\qquad \chi_N&=-\frac{\mathcal N\Sigma}{\mathcal W},\\
 A_\phi&=\frac{q\big[xQ\mathcal N-ar\Dx\mathcal U\big]}{\mathcal W},
 &\qquad \widetilde A_\phi&=-\frac{q\big[xQ\mathcal U+ar\Dx\mathcal N\big]}{\mathcal W}.
 \end{aligned}
 \label{eq:KNLCclosed}
\end{equation}
Here $\widetilde A_\phi=\operatorname{Im}(\Phi_0/\Ecal_0)$ is the magnetic
dual potential; this notation avoids confusion with the Kalb--Ramond field of
Section~\ref{sec:KSLC}.
The dragging function and the temporal gauge potential follow from
\begin{align}
 \partial_r\omega&=\frac{\Dx}{f_N^{2}}
 \left[\partial_x\chi_N+2\operatorname{Im}
 (\bar\Phi_N\partial_x\Phi_N)\right],
 &
 \partial_x\omega&=-\frac{\Dr}{f_N^{2}}
 \left[\partial_r\chi_N+2\operatorname{Im}
 (\bar\Phi_N\partial_r\Phi_N)\right],
 \label{eq:omega-quad}\\[2pt]
 \partial_r A_t&=\frac{\Dx}{f_N}\partial_x\widetilde A_\phi
 -\omega\partial_rA_\phi,
 &
 \partial_x A_t&=-\frac{\Dr}{f_N}\partial_r\widetilde A_\phi
 -\omega\partial_xA_\phi.
 \label{eq:At-quad}
\end{align}
Both integrability conditions vanish identically.  We choose the additive
constants generated by the indefinite integrations to vanish; this fixes the
displayed Killing frame and electromagnetic gauge and gives
\begin{equation}
 \omega_N=-\frac{aP_\omega(r,x)}{N},
 \qquad
 A_t=\frac{\Sigma\mathcal P_A(r,x)}{\mathcal W}.
 \label{eq:omegaAtclosed}
\end{equation}
The polynomial $P_\omega$ is displayed in Appendix~\ref{app:polys}.
$\mathcal P_A$ is odd in $q$ and homogeneous of degree eight in
$(r,m,a,q)$; it is determined uniquely by the quadratures
\eqref{eq:At-quad} together with this additive-constant convention.  Its
expanded form is not printed because it adds no structure to the metric.
Direct substitution verifies both members of each quadrature.
The Einstein--Maxwell solution is therefore
\begin{equation}
 \dd s^{2}=f_N(\dd\phi-\omega_N\dd t)^{2}
 -\frac{\Dr\Dx}{f_N}\dd t^{2}
 +\frac{\mathcal W}{\Sigma}
 \left(\frac{\dd r^{2}}{\Dr}+\frac{\dd x^{2}}{\Dx}\right),
 \qquad
 A=A_t\dd t+A_\phi\dd\phi.
 \label{eq:KNLCmetric}
\end{equation}
With $F=\dd A$, the normalization used here obeys
\begin{equation}
 R_{\mu\nu}=2F_{\mu\lambda}F_\nu{}^{\lambda}
 -\frac12g_{\mu\nu}F_{\alpha\beta}F^{\alpha\beta},
 \qquad
 \nabla_\mu F^{\mu\nu}=0,
 \label{eq:EMeom}
\end{equation}
and hence $R=0$.  The determinant is
$\det g=-(\mathcal W/\Sigma)^{2}$.
At $q=0$ the fields reduce to vacuum KLC, up to the additive constant in
$\omega_N$ recorded in Appendix~\ref{app:polys}.  At $a=0$ they yield the
static Reissner--Nordstr\"om--Levi-Civita member.  The local metric is related to the charged LC representative exhibited in
Ref.~\cite{Astorino:2026backgrounds} through a strong-field limit of
Kerr--Newman--Melvin.  We do not assume global equivalence of the quotient
spacetimes, because the limiting construction and the direct inversion may
inherit different azimuthal identifications.  Equations
\eqref{eq:KNLCclosed}--\eqref{eq:KNLCmetric} provide its direct magnetic-Ernst
representation, including the explicit electromagnetic potentials and the
quadrature proof.
\subsection{Static Reissner--Nordstr\"om--Levi-Civita member}
\label{sec:staticKNLC}
The static $a\to0$ sector is sufficiently compact to display separately.  Set
\begin{equation}
 \Dr=r^{2}-2mr+q^{2},
 \qquad
 \mathfrak D(r,x)=r^{2}\Delta_x+q^{2}x^{2}.
 \label{eq:staticD}
\end{equation}
Then $\omega_N=A_\phi=0$, and the local metric and electromagnetic potentials
reduce to
\begin{align}
 \dd s^{2}={}&-\frac{\Dr\mathfrak D^{2}}{r^{2}}\dd t^{2}
 +\frac{r^{2}\mathfrak D^{2}}{\Dr}\dd r^{2}
 +\frac{r^{2}\mathfrak D^{2}}{\Delta_x}\dd x^{2}
 +\frac{\Delta_x r^{2}}{\mathfrak D^{2}}\dd\phi^{2},
 \label{eq:staticKNLCmetric}\\
 A={}&q\left(m-r-\frac{x^{2}\Dr}{r}\right)\dd t,
 \qquad
 \widetilde A_\phi=-\frac{qx}{\mathfrak D}.
 \label{eq:staticKNLCpotentials}
\end{align}
The additive gauge constant in $A_t$ has been chosen so that $A_t(m,0)=0$.
Direct differentiation gives
\begin{equation}
 \partial_r A_t=\frac{\Delta_x}{f_N}\partial_x\widetilde A_\phi,
 \qquad
 \partial_x A_t=-\frac{\Dr}{f_N}\partial_r\widetilde A_\phi,
 \qquad
 f_N=\frac{\Delta_x r^{2}}{\mathfrak D^{2}},
 \label{eq:staticquadratures}
\end{equation}
so the two electric-potential quadratures are retained exactly in the limit.
The determinant is $\det g=-r^{4}\mathfrak D^{4}$.  As an independent
check beyond the transformation argument, direct symbolic evaluation of the
full four-dimensional curvature and Maxwell divergence verifies all sixteen
Einstein--Maxwell equations and all four Maxwell equations identically for
general $m$ and $q$ in this static sector.
For $q\ne0$ and $r>0$, $\mathfrak D$ is strictly positive on
$-1\le x\le1$.  Thus the static charged member has no Ernst-denominator zero
in this domain.  Its local horizons are
$r_\pm=m\pm\sqrt{m^{2}-q^{2}}$, and the potential pulled back to either
horizon is constant,
\begin{equation}
 \left.A_t\right|_{r=r_\pm}=q(m-r_\pm).
 \label{eq:staticHpotential}
\end{equation}
The area with arbitrary azimuthal period is
$\mathcal A=2\Delta\phi\,r_+^{2}$.
The global limit is more delicate.  Equation~\eqref{eq:conicity-factors}
becomes $\Ccal_N=q^{4}$, so the regular period of the algebraic angle is
\begin{equation}
 \Delta\phi_{\rm reg}=2\pi q^{4},
 \qquad
 \mathcal A_{\rm reg}=4\pi q^{4}r_+^{2}.
 \label{eq:staticperiod}
\end{equation}
The flux formula \eqref{eq:electric-charge} reduces in this sector to
$\mathcal Q_{\rm e}=\Delta\phi/(2\pi q)$; hence
$\mathcal Q_{\rm e}^{\rm reg}=q^{3}$, while the magnetic flux vanishes.
Consequently the local neutral limit of
\eqref{eq:staticKNLCmetric} and the neutral limit of the regular-axis quotient
do not commute: at fixed algebraic $\phi$ the fields tend to the vacuum SLC
member away from the axis, whereas the period in
\eqref{eq:staticperiod} collapses as $q\to0$.  A separately normalized vacuum
SLC quotient must therefore be introduced before comparing global quantities.
This is the Einstein--Maxwell analogue of the noncommuting static/quotient
limit encountered below for KSLC.
\section{Heterotic branch}
\label{sec:KSLC}
\subsection{Action, conventions, and Hassan--Sen map}
In the string frame we use
\begin{equation}
 S=\int\dd^{4}x\sqrt{-g}\,e^{-\varphi}
 \left[R+(\nabla\varphi)^{2}
 -\frac18F_{\mu\nu}F^{\mu\nu}
 -\frac1{12}H_{\mu\nu\rho}H^{\mu\nu\rho}\right].
 \label{eq:heterotic-action}
\end{equation}
Throughout the paper $F=\dd A$ is the conventional exterior derivative.  The
three-form is therefore
\begin{equation}
 H_{\mu\nu\rho}=\partial_\mu B_{\nu\rho}
 +\partial_\nu B_{\rho\mu}+\partial_\rho B_{\mu\nu}
 -\frac14\left(A_\mu F_{\nu\rho}+A_\nu F_{\rho\mu}
 +A_\rho F_{\mu\nu}\right).
 \label{eq:H}
\end{equation}
For reference, the equations checked below can be written as
\begin{align}
 G_{\mu\nu}+\nabla_\mu\nabla_\nu\varphi
 &+\frac12g_{\mu\nu}\big[(\nabla\varphi)^2-2\nabla^2\varphi\big]
 \nonumber\\
 &=\frac14\left\{F_{\mu\lambda}F_\nu{}^\lambda
 +H_{\mu\lambda\rho}H_\nu{}^{\lambda\rho}
 -\frac12g_{\mu\nu}\left(\frac12F^2+\frac13H^2\right)\right\},
 \label{eq:heterotic-Einstein}\\
 \nabla_\mu\!\left(e^{-\varphi}H^{\mu\nu\rho}\right)&=0,
 \label{eq:heterotic-B}\\
 \nabla_\mu\!\left(e^{-\varphi}F^{\mu\nu}\right)
 &=\frac12e^{-\varphi}F_{\alpha\beta}H^{\nu\alpha\beta},
 \label{eq:heterotic-Maxwell}\\
 (\nabla\varphi)^2-2\nabla^2\varphi
 &=R-\frac18F^2-\frac1{12}H^2.
 \label{eq:heterotic-dilaton}
\end{align}
For a stationary vacuum seed $\widetilde g_{\mu\nu}$, the Hassan--Sen map gives 
\begin{align}
 \dd s^{2}_{\rm S}&=\frac{\widetilde g_{tt}}{\Lambda^{2}}
 \left[\dd t+(1+s^{2})\frac{\widetilde g_{t\phi}}
 {\widetilde g_{tt}}\dd\phi\right]^{2}
 +\left(\widetilde g_{\phi\phi}
 -\frac{\widetilde g_{t\phi}^{2}}{\widetilde g_{tt}}\right)\dd\phi^{2}
 +\widetilde g_{rr}\dd r^{2}+\widetilde g_{xx}\dd x^{2},
 \label{eq:HSmetric}\\
 A&=\frac{2s\sqrt{1+s^{2}}}{\Lambda}
 \left[(1+\widetilde g_{tt})\dd t
 +\widetilde g_{t\phi}\dd\phi\right],
 \qquad
 B_{t\phi}=\frac{s^{2}\widetilde g_{t\phi}}{\Lambda},
 \qquad
 \varphi=-\ln\Lambda,
 \label{eq:HSfields}
\end{align}
where
\begin{equation}
 \Lambda=1+s^{2}(1+\widetilde g_{tt}).
 \label{eq:Lambda}
\end{equation}
The Einstein-frame metric is
$g^{\rm E}_{\mu\nu}=e^{-\varphi}g^{\rm S}_{\mu\nu}=\Lambda g^{\rm S}_{\mu\nu}$.
The stationary heterotic sector also admits a larger
Ehlers--Harrison-type symmetry structure \cite{Galtsov1994}; here we retain
the Hassan--Sen route because it maps the vacuum KLC seed directly into the
normalization of \eqref{eq:heterotic-action}.  A real dilaton requires
$\Lambda>0$; this domain restriction will become central in
Section~\ref{sec:KSLCwall}.
Taking the vacuum KLC metric \eqref{eq:KLCcomponents} as the seed gives \cite{Siahaan:2018qcw,Siahaan:2024ljt}
\begin{equation}
 \begin{aligned}
 g^{\rm S}_{tt}&=\frac{\widetilde g_{tt}}{\Lambda^{2}},
 &\qquad g^{\rm S}_{t\phi}&=\frac{(1+s^{2})\widetilde g_{t\phi}}{\Lambda^{2}},\\
 g^{\rm S}_{\phi\phi}&=\widetilde g_{\phi\phi}
 -\frac{s^{2}\bigl(2+2s^{2}+s^{2}\widetilde g_{tt}\bigr)}{\Lambda^{2}}
 \widetilde g_{t\phi}^{2},
 &\qquad g^{\rm S}_{rr}&=\widetilde g_{rr},\\
 && g^{\rm S}_{xx}&=\widetilde g_{xx}.
 \end{aligned}
 \label{eq:KSLCcomponents}
\end{equation}
This division-free form is equivalent to the Hassan--Sen perfect square and
is manifestly regular on a regular seed ergosurface.  Using
$\widetilde g_{tt}\widetilde g_{\phi\phi}-\widetilde g_{t\phi}^{2}
=-\Dr\Dx$, it can equivalently be written with numerator
$(1+s^{2})^{2}\widetilde g_{t\phi}^{2}-\Dr\Dx\Lambda^{2}$ divided by
$\widetilde g_{tt}\Lambda^{2}$.  The factor $(1+s^{2})^{2}$ is fixed by the
perfect square and by the field-equation verification.
For explicit calculations define, with the vacuum $\Dr$ and $N$ in (\ref{eq:vac-blocks}),
\begin{equation}
 \mathcal U_0=\Dx N,
 \qquad
 \mathcal N_0=2amx\big[(x^{2}-3)r^{2}-a^{2}(1+x^{2})\big],
 \qquad
 \mathcal W_0=\mathcal U_0^{2}+\mathcal N_0^{2}.
 \label{eq:vacUNW}
\end{equation}
Then
\begin{equation}
 f_N=\frac{\Dx N\Sigma}{\mathcal W_0},
 \qquad
 \omega_N=-\frac{2mG(r,x)}{a^{3}N},
 \qquad
 \widetilde g_{rr}=\frac{\mathcal W_0}{\Sigma\Dr},
 \qquad
 \widetilde g_{xx}=\frac{\mathcal W_0}{\Sigma\Dx},
 \label{eq:KLCexplicit}
\end{equation}
with $G$ given in Appendix~\ref{app:polys}.  The Killing-block determinant is
\begin{equation}
 g^{\rm S}_{tt}g^{\rm S}_{\phi\phi}
 -(g^{\rm S}_{t\phi})^{2}
 =-\frac{\Dr\Dx}{\Lambda^{2}}.
 \label{eq:detHS}
\end{equation}
Thus the Hassan--Sen parameter never enters the seed horizon polynomial.
At $s=0$ the image reduces to vacuum KLC.  The local $a\to0$ limit, taken
before imposing the rotating-family azimuthal quotient, reproduces the
charged SLC solution of Ref.~\cite{Mazharimousavi:2026xbp}.  The globally
axis-regular rotating quotient has $\Delta\phi\propto a^{2}$, so that quotient
and the static limit need not commute.  In an asymptotically flat Kerr control
calculation the usual parameter identification is
$M=m(1+s^{2})$ and $J=ma(1+s^{2})$ in the present normalization.  No ADM end
exists here, so we do not promote these control relations to global charges of
KSLC.
\section{Local Killing horizons and azimuthal normalization}
\label{sec:horizons}
\subsection{Horizon loci and angular velocities}
For an orthogonally transitive stationary and axisymmetric metric,
the determinant of the Killing block supplies the natural candidate Killing
horizon loci \cite{Carter1969}.  The two determinant identities give
\begin{align}
 \text{KNLC:}\quad&r_\pm=m\pm\delta_N,
 &\delta_N&=\sqrt{m^{2}-a^{2}-q^{2}},
 \label{eq:KNLChorizon}\\
 \text{KSLC:}\quad&r_\pm=m\pm\delta_S,
 &\delta_S&=\sqrt{m^{2}-a^{2}}.
 \label{eq:KSLChorizon}
\end{align}
The corresponding subextremality conditions are
$m^{2}\ge a^{2}+q^{2}$ and $m^{2}\ge a^{2}$.
For KNLC, the second equation in \eqref{eq:omega-quad} contains an overall
factor $\Dr$, so $\partial_x\omega_N=0$ on either horizon.  Therefore
\begin{equation}
 \Omega_H^{N}=\omega_N(r_+)
 \label{eq:OmegaN}
\end{equation}
is constant and $\partial_t+\Omega_H^N\partial_\phi$ is null at $r=r_+$.
The value is shifted by a rigid rotation of the Killing frame, as expected in
a spacetime with no preferred asymptotic rest frame.
For KSLC, \eqref{eq:KLCidentities} and \eqref{eq:KSLCcomponents} give the exact
matched-frame relation
\begin{equation}
 \Omega_H^{S}
 =-\left.\frac{g^{\rm S}_{t\phi}}{g^{\rm S}_{\phi\phi}}\right|_{r_+}
 =\frac{\Omega_H^{\rm KLC}}{1+s^{2}}.
 \label{eq:OmegaKSLC}
\end{equation}
It is unchanged by the string--Einstein conformal transformation.  If the
regular angle is $\phi=\Ccal\varphi$, then
$\Omega_H^{(\varphi)}=\Omega_H^{(\phi)}/\Ccal$; the ratio in
\eqref{eq:OmegaKSLC} remains unchanged when the seed and image use the same
azimuthal normalization.
\subsection{Axis, areas, and surface gravities}
A direct expansion about $x=\pm1$, with $R$ the proper distance from the
axis as in \eqref{eq:KLCconicity}, gives the standard local conicity
test, whose global normalization must be handled before assigning horizon
areas or charges \cite{Astorino2022Conical},
\begin{equation}
 \dd s_{\perp}^{2}=\dd R^{2}+\frac{R^{2}}{\Ccal^{2}}\dd\phi^{2}+O(R^{4}).
 \label{eq:generic-conicity}
\end{equation}
For each family the north and south axes have the same factor.  Denoting the
KNLC and KSLC factors by $\Ccal_N$ and $\Ccal_S$, respectively, one finds
\begin{equation}
 \Ccal_N=q^{4}+16a^{2}m^{2},
 \qquad
 \Ccal_S=\Ccal_0=16a^{2}m^{2}.
 \label{eq:conicity-factors}
\end{equation}
The KSLC result holds in both frames because a regular conformal factor
multiplies the two transverse directions equally.  Hence
\begin{equation}
 \Delta\phi_{\rm reg}^{N}=2\pi\Ccal_N,
 \qquad
 \Delta\phi_{\rm reg}^{S}=2\pi\Ccal_0.
 \label{eq:regular-periods}
\end{equation}
These factors are essential: setting $\Delta\phi=2\pi$ and simultaneously
calling an axis regular would mix different global spacetimes.
The KNLC electromagnetic potentials also determine the conserved fluxes
without reference to an asymptotically flat end.  Their exact axis values are
\begin{equation}
 \left.A_\phi\right|_{x=\pm1}=-\frac{4amq}{\Ccal_N},
 \qquad
 \left.\widetilde A_\phi\right|_{x=\pm1}
 =\mp\frac{q^{3}}{\Ccal_N}.
 \label{eq:axis-potentials}
\end{equation}
For any smooth two-surface homologous to a constant-$(t,r)$ section in a
source-free domain, the conventional Maxwell fluxes are therefore
\begin{align}
 \mathcal Q_{\rm m}
 &=\frac{1}{4\pi}\int_{\mathcal S}F
 =\frac{\Delta\phi}{4\pi}
 \big[A_\phi(1)-A_\phi(-1)\big]=0,
 \label{eq:magnetic-charge}\\
 \mathcal Q_{\rm e}
 &=\frac{1}{4\pi}\int_{\mathcal S}\star F
 =\frac{\Delta\phi}{4\pi}
 \big[\widetilde A_\phi(-1)-\widetilde A_\phi(1)\big]
 =\frac{\Delta\phi}{2\pi}\frac{q^{3}}{\Ccal_N}.
 \label{eq:electric-charge}
\end{align}
On the regular KNLC quotient this becomes the particularly simple result
\begin{equation}
 \mathcal Q_{\rm e}^{\rm reg}=q^{3},
 \qquad
 \mathcal Q_{\rm m}^{\rm reg}=0.
 \label{eq:regular-flux-charge}
\end{equation}
Thus $q$ is a seed parameter rather than the globally normalized electric
charge; the two coincide only after a nonlinear reparametrization on the
regular quotient.
Let $Q_+=r_+^{2}+a^{2}$.  For KNLC,
\begin{equation}
 g_{xx}g_{\phi\phi}=N,
 \qquad
 \left.\sqrt{g_{xx}g_{\phi\phi}}\right|_{r_+}=Q_+.
 \label{eq:KNLCareaelement}
\end{equation}
For KSLC the Einstein-frame area element is
$(1+s^{2})Q_+$, whereas the string-frame element depends on $x$.  With an
arbitrary period $\Delta\phi$ the areas are therefore
\begin{equation}
 \mathcal A_N=2\Delta\phi\,Q_+,
 \qquad
 \mathcal A_{S,E}=2\Delta\phi\,(1+s^{2})Q_+.
 \label{eq:areas-period}
\end{equation}
For the regular quotients,
\begin{equation}
 \mathcal A_N^{\rm reg}=4\pi\Ccal_N Q_+,
 \qquad
 \mathcal A_{S,E}^{\rm reg}=4\pi\Ccal_0(1+s^{2})Q_+.
 \label{eq:areas-regular}
\end{equation}
The second is the local Kerr--Sen area law multiplied by the LC azimuthal
period factor.
In the coordinate normalization of $t$ used in Sections~\ref{sec:KNLC} and
\ref{sec:KSLC}, direct evaluation gives
\begin{equation}
 \kappa_N=\frac{\delta_N}{Q_+},
 \qquad
 \kappa_S=\frac{\delta_S}{(1+s^{2})Q_+}.
 \label{eq:kappas}
\end{equation}
The KSLC value is the same in string and Einstein frames because $\Lambda$ is
regular on the horizon and invariant along its generator.  A constant
rescaling of $t$ rescales both $\kappa$ and $\Omega_H$; without an asymptotic
unit timelike Killing field there is no canonical normalization.
It is useful to remove the purely global azimuthal period from the entropy by
defining
\begin{equation}
 \widehat S\equiv\frac{2\pi}{\Delta\phi}\frac{\mathcal A}{4}.
 \label{eq:normalized-entropy}
\end{equation}
Then
\begin{equation}
 \widehat S_N=\pi Q_+,
 \qquad
 \widehat S_{S,E}=\pi(1+s^{2})Q_+,
 \qquad
 2T_N\widehat S_N=\delta_N,
 \qquad
 2T_S\widehat S_{S,E}=\delta_S,
 \label{eq:TSidentity}
\end{equation}
where $T=\kappa/(2\pi)$.  These are local horizon identities, not a claim of a
global first law.  The KNLC Maxwell flux is fixed by
\eqref{eq:electric-charge}, but such a law would still require a global mass
and angular momentum, a reference-fixed horizon electric potential, the
appropriate heterotic gauge charges, and boundary terms for the chosen LC
completion.
\section{Singularities, causal structure, and algebraic type}
\label{sec:singularities}
\subsection{Exact factorization and the former Kerr ring in KNLC}
\label{sec:ring}
The polynomial $\mathcal W$ in \eqref{eq:UNW} has the exact factorization
\begin{equation}
 \mathcal W=\Sigma\mathcal H,
 \label{eq:WHfactor}
\end{equation}
where the degree-six polynomial $\mathcal H$ is given in
Appendix~\ref{app:polys}.  Consequently
\begin{equation}
 f_N=\frac{\Dx N}{\mathcal H},
 \qquad
 \chi_N=-\frac{\mathcal N}{\mathcal H},
 \qquad
 g_{rr}=\frac{\mathcal H}{\Dr},
 \qquad
 g_{xx}=\frac{\mathcal H}{\Dx},
 \qquad
 \det g=-\mathcal H^{2}.
 \label{eq:cancelled-metric}
\end{equation}
The numerators of both $A_\phi$ and $\widetilde A_\phi$ are separately
divisible by $\Sigma$, and $A_t=\mathcal P_A/\mathcal H$.  Thus every metric
and gauge component has a finite rational continuation through the former
Kerr ring $r=0=x$ whenever $m\,a\,q\neq0$.
At that locus,
\begin{equation}
 \mathcal H_*=4a^{4}m^{2},
 \qquad
 N_*=-a^{2}q^{2},
 \qquad
 f_{N*}=-\frac{q^{2}}{4a^{2}m^{2}},
 \label{eq:ring-values1}
\end{equation}
while
\begin{equation}
 \begin{aligned}
 (g_{rr})_*&=\frac{4a^{4}m^{2}}{a^{2}+q^{2}},
 &\qquad (g_{xx})_*&=4a^{4}m^{2},\\
 \omega_*&=\frac{a(4a^{2}m^{2}+a^{2}q^{2}+q^{4})}{q^{2}},
 &\qquad (\det g)_*&=-16a^{8}m^{4}.
 \end{aligned}
 \label{eq:ring-values2}
\end{equation}
The metric is analytic and nondegenerate in this chart, so all polynomial
curvature invariants are finite there.  Electromagnetic charge therefore does
not restore the Kerr curvature ring.
It does, however, change its causal character.  Since
$f_N=g_{\phi\phi}<0$ at the ring, the periodic orbits of $\partial_\phi$ are
timelike in an open neighbourhood.  Generic charged KNLC therefore contains
an \emph{interior} azimuthal closed-timelike-curve region.  In the subextreme
case $r=0$ lies behind the inner horizon, so this does not contradict the
exterior result below, but it shows that curvature regularity is not equivalent
to causal regularity.
\subsection{No exterior Ernst zeros or azimuthal closed timelike curves}
\label{sec:noexteriorctc}
\begin{quote}
\textbf{Proposition 2.}
Let $m>0$, $q\ne0$, and $m^{2}\ge a^{2}+q^{2}$.  In the domain
$r\ge r_+$, $|x|\le1$, one has $\mathcal W>0$.  Moreover
$g_{\phi\phi}>0$ away from the axis and $g_{\phi\phi}=0$ only on the axis.
Thus the KNLC exterior contains neither an Ernst-zero singularity nor an
azimuthal closed timelike curve.
\end{quote}
\noindent
For $r\ge r_+$, $\Dr\ge0$ and
\begin{align}
 N&=Q^{2}-a^{2}\Dr\Dx
 \ge Q^{2}-a^{2}\Dr
 \nonumber\\
 &=r^{4}+a^{2}r^{2}+2a^{2}mr-a^{2}q^{2}
 \ge r^{4}+a^{2}r^{2}+a^{2}mr>0,
 \label{eq:Npositive}
\end{align}
where $q^{2}\le m^{2}\le mr$ was used in the last step.  If $|x|<1$, then
$\mathcal U=\Dx N+q^{2}(r^{2}x^{2}+a^{2})>0$.  On the axis,
$\mathcal U=q^{2}(r^{2}+a^{2})>0$.  Hence
$\mathcal W=\mathcal U^{2}+\mathcal N^{2}>0$.  Since $\Sigma>0$ in the
exterior, $\mathcal H>0$, and
$f_N=\Dx N/\mathcal H$ has the stated sign.  This proof excludes closed curves
generated by the axial Killing field; it is not a classification of all
possible non-azimuthal causal pathologies in a maximal extension.
\subsection{The finite-radius heterotic wall}
\label{sec:KSLCwall}
The Hassan--Sen factor cannot remain positive throughout a full LC end.  At
fixed $|x|<1$, the KLC seed has the exact large-$r$ behaviour
\begin{align}
 r^{2}f_N&\longrightarrow\frac{1}{\Delta_x},
 \nonumber\\
 \frac{\omega_N}{r}&\longrightarrow
 -2am(x^{4}-6x^{2}-3),
 \label{eq:KLCasym1}\\
 \widetilde g_{tt}&=-(1-x^{2})^{2}r^{4}
 +2m(1-x^{2})^{2}r^{3}+O(r^{2}).
 \label{eq:KLCasym2}
\end{align}
A finite rigid shift of $\omega_N$ affects only subleading terms in
\eqref{eq:KLCasym2}.  Therefore, for every $s\ne0$,
\begin{equation}
 \Lambda=1+s^{2}(1+\widetilde g_{tt})
 =-s^{2}(1-x^{2})^{2}r^{4}+O(r^{3})<0
 \label{eq:Lambda-asym}
\end{equation}
for sufficiently large $r$ at any fixed off-axis latitude.
On the outer seed horizon, by contrast,
$\widetilde g_{tt}=f_N\omega_N^{2}\ge0$, and hence
\begin{equation}
 \Lambda|_{r_+}\ge1+s^{2}>0.
 \label{eq:Lambda-horizon}
\end{equation}
Continuity then guarantees at least one root
$r_\Lambda(x)>r_+$ on every fixed ray $|x|<1$.  Let $r_\Lambda(x)$ denote the
first such root reached from the horizon.  Since $e^{-\varphi}=\Lambda$, the
real heterotic branch is restricted to $\Lambda>0$.  At any finite-order root,
\begin{equation}
 (\nabla\varphi)^{2}
 =\frac{(\nabla\Lambda)^{2}}{\Lambda^{2}}
 \label{eq:dilaton-invariant}
\end{equation}
diverges as an invariant scalar.  The metric components in both frames also
contain negative powers of $\Lambda$.  The surface is therefore a genuine
dilaton/matter singularity, not a removable sign convention.
For small $|s|$, the leading location is
\begin{equation}
 r_\Lambda(x)=\frac{1}{\sqrt{|s|}\sqrt{1-x^{2}}}+O(1).
 \label{eq:wall-small-s}
\end{equation}
The wall recedes to infinity as $s\to0$ and also toward the symmetry axis.
The $\Lambda>0$ domain may consequently have narrowing axial channels, a
global question not settled here.  What is settled by
\eqref{eq:Lambda-asym} is that the horizon branch has no regular LC asymptotic
end covering a finite angular interval.  KSLC should therefore be regarded as
a local exact heterotic geometry with a regular Killing horizon, not as a
completed black-hole exterior asymptotic to a ``dilatonic LC universe.''
\subsection{Polynomial curvature invariants}
\label{sec:curvature}
\begin{figure}[!t]
\includegraphics[width=0.98\textwidth]{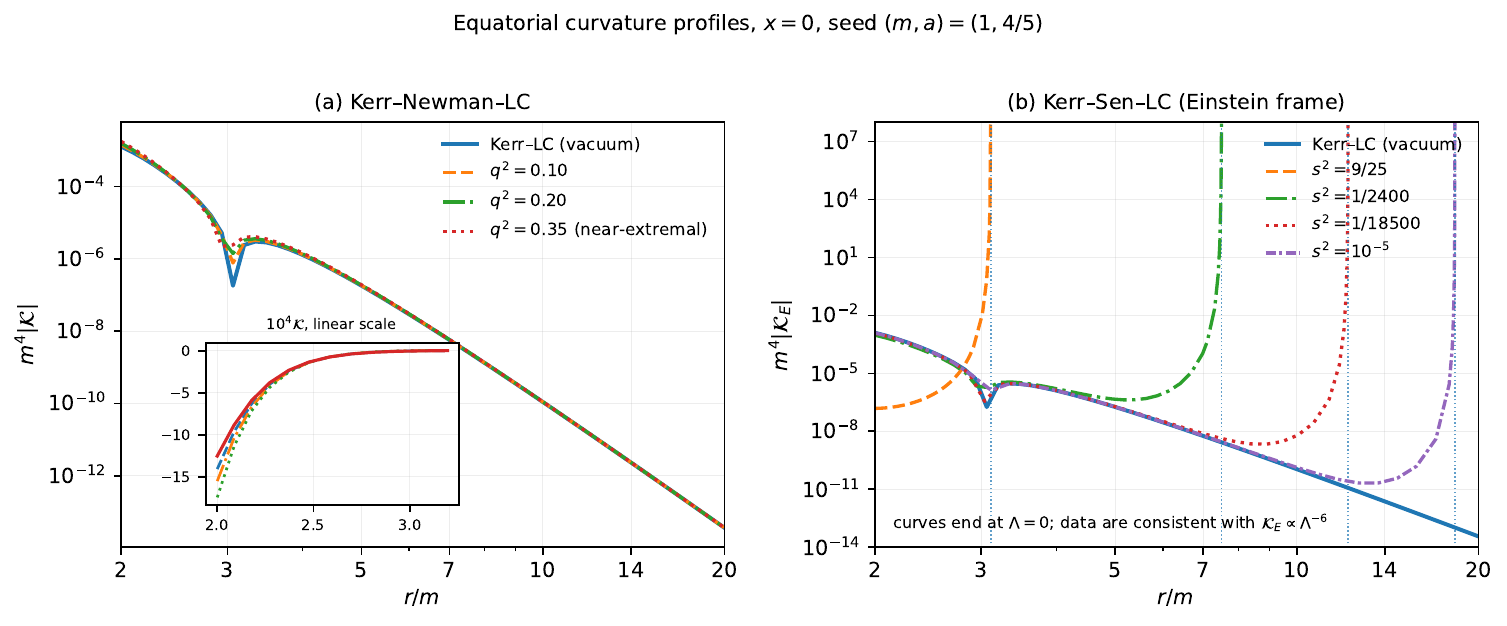}
\caption{Equatorial Kretschmann profiles for charge families at fixed seed
$(m,a)=(1,4/5)$.
(a)~Einstein--Maxwell branch: the KNLC family
$q^{2}=0.10,0.20,0.35$ is nearly degenerate with the vacuum \KLC{} curve on a
logarithmic scale; the inset (linear scale, $\mathcal K\times10^{4}$) resolves the
small charge ordering near the sign change of $\mathcal K$ at
$r\simeq3$.  All curves approach the universal law \eqref{eq:K-universal}.
(b)~Heterotic branch (Einstein frame): members with
$s^{2}=9/25,\,1/2400,\,1/18500,\,10^{-5}$ each terminate at their own
$\Lambda=0$ wall (dotted verticals), where
the Einstein-frame data are numerically consistent with
$\mathcal K^{(E)}\propto\Lambda^{-6}$; the wall recedes as
$r_{w}\sim(s^{2})^{-1/4}$ at small charge, cf.\ \eqref{eq:wall-small-s}.
The curves were evaluated with exact rational arithmetic before conversion
to plotting precision.}
\label{fig:kfamilies}
\end{figure}

\begin{figure}[!t]
\includegraphics[width=0.98\textwidth]{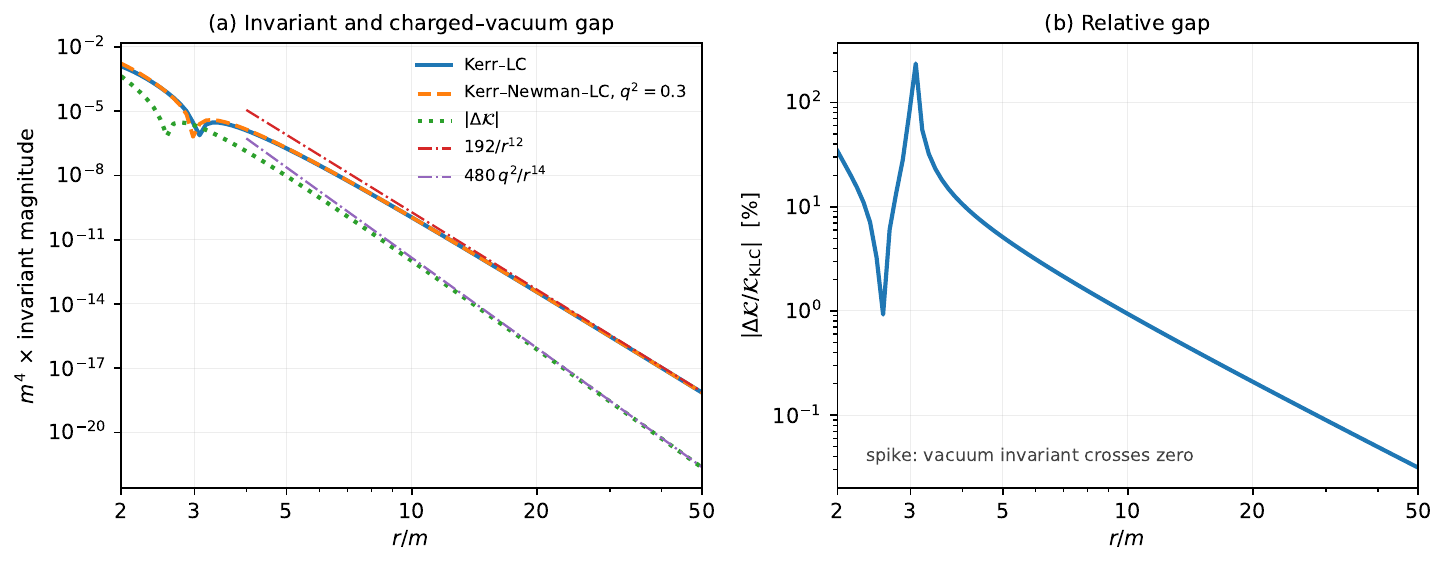}
\caption{The charged--vacuum Kretschmann gap in the Einstein--Maxwell branch
at $x=0$, $(m,a,q^{2})=(1,4/5,0.3)$, evaluated exactly from the verified
symbolic invariant.  (a)~$|\mathcal K|$ for KNLC and \KLC{} together with
$|\Delta\mathcal K|$: the gap decays as $480q^{2}/r^{14}$
\eqref{eq:K-gap}, two powers faster than the invariants themselves
\eqref{eq:K-universal}.  (b)~Relative gap: apart from the spike where
$\mathcal K_{\rm KLC}$ crosses zero ($r\simeq3.03$), the relative difference
falls monotonically ($0.94\%$ at $r=10$, $0.031\%$ at $r=50$).}
\label{fig:kgap}
\end{figure}

The Kretschmann scalar
$\mathcal K\equiv R_{\mu\nu\rho\sigma}R^{\mu\nu\rho\sigma}$ was obtained
symbolically for KNLC with GRTensorIII and evaluated independently by an
exact-rational local-jet Riemann engine.  The two results agree exactly at 30
independent rational points across three parameter sets.  For KSLC,
GRTensorIII supplies the string-frame equatorial and smooth-axis slice
expressions.  The equatorial expression agrees exactly with the jet result at
21 rational points, while the axial expression agrees with the off-axis
$x\to1$ limit to high precision.  Einstein-frame ring values and wall
exponents quoted below are obtained from the independent jet calculation;
the detailed provenance is recorded in Appendix~\ref{app:verification}.
For the Einstein--Maxwell branch the full invariant has the factored form
\begin{equation}
 \mathcal K_{\rm KNLC}
 =\frac{8\,\mathcal P_{\rm N}(r,x)}{\mathcal H(r,x)^{6}},
 \label{eq:K-KNLC-struct}
\end{equation}
where $\mathcal P_{\rm N}$ is a polynomial of degree 24 in each of $r$ and
$x$ whose expanded form is too long to display.  The denominator contains the
same structure polynomial as $\det g=-\mathcal H^{2}$.  Together with
Proposition~2, Eq.~\eqref{eq:K-KNLC-struct} makes the absence of ordinary
polynomial-curvature poles in the subextreme exterior explicit.  Direct
symbolic substitution into the factorized expression gives the finite former
ring value
\begin{equation}
 \mathcal K_{\rm KNLC}(0,0)
 =\frac{7q^{4}\!\left(4q^{2}+3a^{2}\right)^{2}
 +72m^{2}a^{2}\!\left[q^{2}\!\left(2q^{2}+a^{2}\right)
 -m^{2}\!\left(4q^{2}+a^{2}\right)\right]}{32m^{8}a^{12}},
 \label{eq:K-ring-KNLC}
\end{equation}
which reduces to
\begin{equation}
 \mathcal K_{\rm KLC}(0,0)=-\frac{9}{4m^{4}a^{8}}
 \label{eq:K-ring-KLC}
\end{equation}
as $q\to0$.  The Lorentzian contraction is not positive definite, so its
sign should not be interpreted as a curvature norm.  A sign change can occur
inside the black-hole parameter domain; for example, at $(m,a)=(1,1/5)$ the
positive root is $q^{2}\simeq0.41625$, below the subextreme upper bound
$m^{2}-a^{2}=0.96$.  Curvature regularity does not remove the interior
closed-timelike-curve region identified in Sec.~\ref{sec:ring}.
At fixed $|x|<1$, the vacuum and charged LC ends share the exact leading law
\begin{equation}
\mathcal K_{\rm K(N)LC}
 =\frac{192}{{\Delta_x}^{6}r^{12}}+O(r^{-13}),
 \qquad r\to\infty.
 \label{eq:K-universal}
\end{equation}
This expansion is not uniform as $x\to\pm1$.  On the equator, direct symbolic
expansion further gives
\begin{equation}
 \mathcal K_{\rm KNLC}-\mathcal K_{\rm KLC}
 =\frac{480q^{2}}{r^{14}}-\frac{1152mq^{2}}{r^{15}}
 +O(r^{-16}),
 \qquad x=0.
 \label{eq:K-gap}
\end{equation}
Thus charge is absent from the leading far-field curvature and first appears
two inverse powers later on this slice.  Figures~\ref{fig:kfamilies} and
\ref{fig:kgap} illustrate these results.
The heterotic branch behaves differently.  In the string frame the general
expression is too large to display, but the equatorial and smooth-axis limits
have the exact structure
\begin{equation}
 \left.\mathcal K^{(S)}_{\rm KSLC}\right|_{x=x_\star}
 =\frac{4\mathcal P_\star(r)}
 {a^{24}\mathcal H_{0}(r,x_\star)^{10}\Lambda(r,x_\star)^{4}},
 \qquad x_\star=0,1,
 \label{eq:K-KSLC-slices}
\end{equation}
where $x_\star=1$ denotes $\lim_{x\to1}\mathcal K(r,x)$ and
$\mathcal H_0=\mathcal W_0/\Sigma$.  The denominator factorizations are
symbolic identities.  Hence, on these slices, a generic simple wall root for
which the reduced numerator is nonzero is a fourth-order string-frame
curvature pole,
\begin{equation}
 \mathcal K^{(S)}_{\rm KSLC}\propto\Lambda^{-4}.
 \label{eq:K-wall-string}
\end{equation}
At a representative equatorial wall, independent Einstein-frame local-jet
data over three decades are numerically consistent with
\begin{equation}
 \mathcal K^{(E)}_{\rm KSLC}\propto\Lambda^{-6},
 \qquad R^{(E)}\propto\Lambda^{-3}.
 \label{eq:K-wall-Einstein}
\end{equation}
These exponents are obtained from the full conformally transformed metric,
including derivatives of $\Lambda$; they do not follow from a simple
conformal weight.  The wall is therefore a curvature singularity in both
frames at the tested generic roots.
Within the connected physical component $\Lambda>0$, the local horizon and
smooth axial segments with $\Lambda_{\rm ax}\neq0$ have finite ordinary
curvature invariants.  An axial intersection with the wall is singular, and a
finite smooth-axis limit does not include distributional curvature from an
unregularized conical defect.  The former-ring values are finite generically
provided $\Lambda_\ast\equiv\Lambda(0,0)\neq0$.  As an illustrative
comparison, use the asymptotically flat progenitor identifications
$M_{\rm flat}=m_S(1+s^{2})$, $J_{\rm flat}=M_{\rm flat}a$, and
$Q_{\rm flat}^{2}=2m_S^{2}s^{2}(1+s^{2})$.  The KSLC point
$(m_S,a,s^{2})=(1,4/5,9/25)$ then corresponds to the KNLC control parameters
$(m_N,a,q^{2})=(34/25,4/5,612/625)$.  The numerical Einstein-frame KSLC ring
limit is approximately $21.65$, whereas Eq.~\eqref{eq:K-ring-KNLC} gives
approximately $-10.60$.  This is a progenitor-parameter comparison, not a
matching of global LC charges.
\subsection{Petrov classification}
\label{sec:petrov}
The Petrov type was tested independently from the field-equation worksheets.
At each point we formed the electric and magnetic Weyl tensors in an
orthonormal frame and the complex trace-free matrix
$\mathsf Q=\mathsf E+i\mathsf B$.  The associated $I$--$J$ discriminant is
the invariant basis of the speciality-index diagnostic
\cite{BakerCampanelli2000}.  A repeated eigenvalue requires the vanishing of
\begin{equation}
 \mathcal D=\frac12\left[\operatorname{tr}(\mathsf Q^{2})\right]^{3}
 -3\left[\operatorname{tr}(\mathsf Q^{3})\right]^{2}.
 \label{eq:PetrovD}
\end{equation}
We quote the dimensionless diagnostic
\begin{equation}
 \delta_{\rm P}=\frac{|\mathcal D|}{
 \max\left\{\tfrac12|\operatorname{tr}(\mathsf Q^{2})|^{3},
 3|\operatorname{tr}(\mathsf Q^{3})|^{2}\right\}}.
 \label{eq:Petrovdiagnostic}
\end{equation}
A 64-bit local-jet implementation gives
\begin{center}
\begin{tabular}{lcc}
\hline\hline
geometry and point & parameters & $\delta_{\rm P}$\\
\hline
Kerr control, $(r,x)=(23/7,2/5)$
& $(m,a)=(13/10,7/10)$ & $1.5\times10^{-15}$\\
KNLC, $(r,x)=(23/7,2/5)$
& $(m,a,q)=(13/10,7/10,2/5)$ & $1.959\times10^{-1}$\\
KSLC, $(r,x)=(12/5,1/5)$
& $(m,a,s)=(5/4,3/4,3/4)$ & $1.059\times10^{-1}$\\
\hline\hline
\end{tabular}
\end{center}
The Kerr control reproduces the repeated Weyl eigenvalue to floating-point
accuracy, while the two charged LC values are separated from zero by order-one
fractions of the individual terms in \eqref{eq:PetrovD}.  Therefore neither
discriminant vanishes identically, and both analytic families are generically
Petrov type~I.  Algebraically special parameter subloci, including static or
degenerate limits, are not classified here.  The KSLC conclusion is conformal
frame independent.
\section{Comparison and global scope}
\label{sec:scope}
The two principal constructions share a vacuum KLC limit but charge it in
fundamentally different ways.  The main structural facts are summarized in
Table~\ref{tab:comparison}.  Section~\ref{sec:gauge-harrison-order} also
identifies an auxiliary Harrison-dressed KLC branch generated by
$\mathsf H_b\circ\mathsf I=\mathsf I\circ\mathsf D_b$.  Its vacuum horizon
polynomial and regular-quotient flux
$\left.\mathcal Q_{\rm e}^{(b)}\right|_{\rm reg}=4abm$ distinguish its
displayed parametrization from
the KNLC route, but a full invariant comparison allowing parameter remixing,
coordinate changes, electromagnetic duality, and identifications with known
magnetized or swirling families has not been completed.  It is therefore not
counted as a third globally analysed branch in Table~\ref{tab:comparison}.
The local Einstein--Maxwell line element is not independent of the
representative reported in Ref.~\cite{Astorino:2026backgrounds}.  That work
obtains it as a strong-field limit of Kerr--Newman--Melvin and explains that
inversion is a special Harrison/Ehlers operation.  A global identification,
however, requires the azimuthal periods and conical quotients to be matched;
none is assumed here.  The novel content claimed here for that
branch is correspondingly limited to: the direct electrovacuum Ernst proof;
the compact potentials \eqref{eq:KNLCclosed}; the two closed rational
quadratures; the exact field-equation record; the explicit static member
\eqref{eq:staticKNLCmetric}; the charge-dependent conicity and flux charge
\eqref{eq:regular-flux-charge}; the factorization
\eqref{eq:WHfactor}; the analytic ring regularity together with the newly
exposed interior CTC region; the exterior positivity theorem; and the generic
Petrov-I check.  The heterotic Hassan--Sen image and its full
rotating $(A,B,\varphi)$ sector appear, to our knowledge, not to have been
reported previously.
All equations proved in this paper are local equations on the displayed
coordinate domain.  They do not establish geodesic completeness, a unique
maximal extension, or the absence of distributional matter at a boundary of
the chart.  The Weyl-coordinate analysis of Ref.~\cite{HerdeiroNovo2026}
provides a concrete warning: other symmetry-transformed static solutions can
obey the vacuum equations pointwise and nevertheless be supported by a hidden
annular source.  A corresponding canonical-Weyl map, rod analysis, junction
analysis across any degenerate coordinate surface, and study of geodesic
affine completeness remain necessary for both rotating branches.  The
heterotic wall adds a separate obstruction even before that global programme
is attempted.
For the same reason, the local area and surface-gravity formulae of
Section~\ref{sec:horizons} should not yet be promoted to a complete
thermodynamic first law.  Although the KNLC Maxwell flux is fixed by
\eqref{eq:electric-charge}, the LC end supplies no canonical normalization of
$\partial_t$, the azimuthal quotient changes all extensive horizon quantities,
and the KSLC branch lacks a regular full LC infinity.  Quasilocal or
covariant phase-space definitions of mass and angular momentum relative to a
fixed LC reference, a reference-fixed horizon electric potential, and the
work term associated with varying the conicity are natural next steps in the
Noether-charge framework \cite{Wald1993,IyerWald1994}.
Two further continuations are suggested directly by the surrounding
literature.  First, generalized Harrison maps now exist in nonlinear
Einstein--ModMax sectors \cite{BokulicHerdeiro2025}; applying an inversion-like
operation to a rotating nonlinear-electrodynamic seed would test whether the
ring cancellation found here is a special consequence of the quadratic
Maxwell Ernst system.  Second, photon rings and shadows in swirling
backgrounds exhibit characteristic spin--background couplings
\cite{Capobianco2026}.  A corresponding geodesic and optical analysis is
well-posed for the KNLC exterior established in
Section~\ref{sec:noexteriorctc}.  For KSLC, however, any observer/source setup
must remain inside the $\Lambda>0$ component bounded by the finite-radius wall;
this restriction is physical rather than a coordinate choice.
\begin{table}[!t]
\caption{Structural comparison of the two principal charged LC branches analysed globally in this paper.}
\label{tab:comparison}
\small
\centering
\begin{tabular}{lp{5.0cm}p{5.0cm}}
\hline\hline
 & Einstein--Maxwell KNLC & heterotic KSLC\\
\hline
construction & inversion of Kerr--Newman & Hassan--Sen of vacuum KLC\\
$\Dr$ & $r^{2}-2mr+a^{2}+q^{2}$ & $r^{2}-2mr+a^{2}$\\
additional fields & $A_\mu$ & $A_\mu,B_{\mu\nu},\varphi$\\
outer local horizon & Kerr--Newman radius & Kerr radius\\
regular-axis factor & $q^{4}+16a^{2}m^{2}$ & $16a^{2}m^{2}$\\
regular electric flux & $\mathcal Q_{\rm e}=q^{3}$ & not assigned here\\
former Kerr ring & curvature-regular; interior CTC & seed-regular locally\\
ring $\mathcal K$ & finite exact quartic, Eq.~\eqref{eq:K-ring-KNLC} & finite if $\Lambda_\ast\neq0$\\
far-field $\mathcal K$ & $192/[(1-x^{2})^{6}r^{12}]$ & terminates at wall\\
wall divergence & none & exact $\Lambda^{-4}$ string-frame slices; Einstein exponent numerical\\
outer domain & no Ernst zeros; $g_{\phi\phi}>0$ & finite-radius $\Lambda=0$ wall\\
algebraic type & generic I & generic I\\
\hline\hline
\end{tabular}
\end{table}
\section{Conclusions}
\label{sec:conclusion}
We have given a unified construction and analysis of two principal charged
rotating Kerr--Levi-Civita geometries.  In Einstein--Maxwell theory the
magnetic Kerr--Newman Ernst pair can be inverted exactly.  We also established
the exact conjugacy
$\mathsf I\circ\mathsf H_c=\mathsf D_c\circ\mathsf I$, which shows that the
electromagnetic representative of the seed must be fixed before inversion
and prevents the standard Kerr--Newman inversion from being identified with
a bare Harrison-then-inversion composition.  The conjugate ordering produces
an auxiliary Harrison-dressed KLC branch with the vacuum Kerr polynomial and,
on its regular quotient, electric flux
$\left.\mathcal Q_{\rm e}^{(b)}\right|_{\rm reg}=4abm$; its full invariant
classification remains
open.  For the principal KNLC branch, the transformed dragging function and
temporal potential follow from integrable quadratures and the full solution
is rational.  This direct representation complements Astorino's independent
strong-field construction of the same local KNLC line element; the global
azimuthal quotients are not identified without a separate matching.  Its $a=0$ member
has the compact form
\eqref{eq:staticKNLCmetric}; for $q\ne0$ its denominator is positive for
$r>0$, its horizon electrostatic potential is constant, and its regular-axis
period is $2\pi q^{4}$.  More generally, the KNLC axis data give zero magnetic
flux and the exact regular-quotient electric charge
$\mathcal Q_{\rm e}^{\rm reg}=q^{3}$.  The neutral local limit and the
neutral limit of the regular quotient are therefore inequivalent.
The Einstein--Maxwell branch provides a sharp answer to the original
regularization question.  Its common Ernst denominator factorizes by the Kerr
ring factor, all metric and gauge components are analytic at $r=x=0$, and the
polynomial curvature singularity remains absent for generic $m\,a\,q\neq0$.
Charge nevertheless creates an interior region with timelike azimuthal
orbits.  Outside the outer horizon the situation is cleaner: an analytic
positivity argument excludes both Ernst zeros and azimuthal closed timelike
curves.
The heterotic branch gives the opposite global lesson.  Hassan--Sen charging
preserves the Kerr horizon polynomial and yields a regular local Killing
horizon with the expected $(1+s^{2})$ rescaling of angular velocity, area, and
surface gravity.  Yet $\Lambda$ becomes negative at large radius in every
fixed off-axis direction.  The real branch therefore terminates at a singular
$\Lambda=0$ wall and does not define a standard LC-asymptotic black-hole
exterior.  This obstruction is invisible in a local zero check of the field
equations and is the most important qualification of the heterotic
construction.
The curvature analysis makes both lessons quantitative.  The full KNLC
Kretschmann scalar is $8\mathcal P_{\rm N}/\mathcal H^{6}$ and is manifestly
pole-free in the subextreme exterior.  Direct symbolic evaluation gives the
finite former-ring value \eqref{eq:K-ring-KNLC}, while at fixed off-axis
latitude the far field obeys the universal law
$\mathcal K\sim192/[(1-x^{2})^{6}r^{12}]$.  For KSLC, exact string-frame
factorizations establish a generic $\Lambda^{-4}$ pole on the equatorial and
smooth-axis slices.  Independent Einstein-frame data at a representative
equatorial wall are numerically consistent with
$\mathcal K^{(E)}\propto\Lambda^{-6}$ and $R^{(E)}\propto\Lambda^{-3}$.
The horizon and smooth axial segments inside the first wall have finite
ordinary invariants, and the former ring is finite when
$\Lambda_\ast\neq0$.  The verification record distinguishes symbolic
identities from exact rational point tests and high-precision limiting
calculations.
Both families are generically Petrov type~I.  Their regular-axis periods and
horizon areas differ by charge-dependent global factors, so future work on
thermodynamics must keep the azimuthal quotient explicit.  The highest-priority
open problems are a canonical-Weyl and rod analysis of the maximal extension,
a search for distributional sources, the topology and causal role of the
heterotic wall and its axial channels, and covariant definitions of mass,
angular momentum, heterotic gauge charges, and a first law in a fixed
Levi-Civita environment.  The explicit static
member, nonlinear-electrodynamic generalizations, and optical observables of
the regular KNLC exterior provide three concrete follow-up projects.
\appendix
\section{Proof that inversion preserves the electrovacuum Ernst system}
\label{app:proof}
Throughout this appendix the dot is the symmetric complex-bilinear product on
the Weyl base.  Let $(\Ecal_0,\Phi_0)$ solve \eqref{eq:ernsteqs}, let
$f_0$ and $\bm J_0$ be defined by \eqref{eq:ernstdefs}, and set
\begin{equation}
 \Ecal=\Ecal_0^{-1},
 \qquad
 \Phi=\Phi_0\Ecal_0^{-1}.
 \label{eq:app-map}
\end{equation}
The phase $e^{i\alpha}$ can be restored at the end.
First,
\begin{equation}
 f=\operatorname{Re}\frac{1}{\Ecal_0}
 +\varepsilon\left|\frac{\Phi_0}{\Ecal_0}\right|^{2}
 =\frac{\operatorname{Re}\Ecal_0+\varepsilon|\Phi_0|^{2}}
 {|\Ecal_0|^{2}}
 =\frac{f_0}{|\Ecal_0|^{2}}.
 \label{eq:app-f}
\end{equation}
Next, using
$\nabla\Ecal=-\Ecal_0^{-2}\nabla\Ecal_0$ and
$\nabla\Phi=\Ecal_0^{-1}\nabla\Phi_0
-\Phi_0\Ecal_0^{-2}\nabla\Ecal_0$, one obtains
\begin{equation}
 \bm J=\Ecal_0^{-2}\bar\Ecal_0^{-1}
 \left[-(\bar\Ecal_0+2\varepsilon|\Phi_0|^{2})\nabla\Ecal_0
 +2\varepsilon\Ecal_0\bar\Phi_0\nabla\Phi_0\right].
 \label{eq:app-Jraw}
\end{equation}
The definition of $f_0$ implies
$\bar\Ecal_0+2\varepsilon|\Phi_0|^{2}=2f_0-\Ecal_0$, and hence
\begin{equation}
 \bm J=\Ecal_0^{-2}\bar\Ecal_0^{-1}
 (\Ecal_0\bm J_0-2f_0\nabla\Ecal_0).
 \label{eq:app-J}
\end{equation}
For the gravitational Ernst equation,
\begin{align}
 f\nabla^{2}\Ecal
 &=\Ecal_0^{-4}\bar\Ecal_0^{-1}
 \left[-\Ecal_0(\bm J_0\mathbin{\cdot}\nabla\Ecal_0)
 +2f_0(\nabla\Ecal_0\mathbin{\cdot}\nabla\Ecal_0)\right],
 \label{eq:app-E-left}\\
 \bm J\mathbin{\cdot}\nabla\Ecal
 &=\Ecal_0^{-4}\bar\Ecal_0^{-1}
 \left[-\Ecal_0(\bm J_0\mathbin{\cdot}\nabla\Ecal_0)
 +2f_0(\nabla\Ecal_0\mathbin{\cdot}\nabla\Ecal_0)\right],
 \label{eq:app-E-right}
\end{align}
where the seed equation was used in the first line.  The two sides coincide.
For the electromagnetic Ernst equation, expansion of
$\nabla^{2}(\Phi_0\Ecal_0^{-1})$ and use of both seed equations gives
\begin{align}
 f\nabla^{2}\Phi
 =\Ecal_0^{-4}\bar\Ecal_0^{-1}
 \big[&\Ecal_0^{2}(\bm J_0\mathbin{\cdot}\nabla\Phi_0)
 -\Ecal_0\Phi_0(\bm J_0\mathbin{\cdot}\nabla\Ecal_0)
 -2f_0\Ecal_0(\nabla\Ecal_0\mathbin{\cdot}\nabla\Phi_0)
 \nonumber\\
 &+2f_0\Phi_0(\nabla\Ecal_0\mathbin{\cdot}\nabla\Ecal_0)\big].
 \label{eq:app-Phi-left}
\end{align}
On the other hand,
\begin{equation}
 \bm J\mathbin{\cdot}\nabla\Phi
 =\Ecal_0^{-4}\bar\Ecal_0^{-1}
 (\Ecal_0\bm J_0-2f_0\nabla\Ecal_0)
 \mathbin{\cdot}
 (\Ecal_0\nabla\Phi_0-\Phi_0\nabla\Ecal_0),
 \label{eq:app-Phi-right}
\end{equation}
whose expansion is exactly \eqref{eq:app-Phi-left}.  Thus both transformed
Ernst equations hold for either $\varepsilon=+1$ or $-1$.
Finally, a constant factor $c$ multiplying $\Phi_0/\Ecal_0$ enters only as
$c\bar c$ in $f$ and $\bar\Phi\nabla\Phi$.  The proof therefore remains valid
for $|c|=1$, giving the phase in \eqref{eq:inversion-full}; for $|c|\ne1$ the
residual is proportional to $|c|^{2}-1$.
\section{Verification record and convention audit}
\label{app:verification}
Every displayed claim was checked at three independent levels.
(i)~\emph{Computer algebra.}  GRTensorII/III worksheets construct both
branches and reduce the field equations to zero componentwise.
(ii)~\emph{Independent exact-rational evaluation.}  A second-order automatic
differentiation over the rationals represents every field by its exact 2-jet
at rational points and contracts the Christoffel and Riemann data without
expanding any large expression, so a vanishing residual is exact rather than
a small floating-point number.  This engine confirms all $16+4$ (KNLC) and
$16+6+4+1$ (KSLC) field-equation residuals at several generic rational
parameter and coordinate points; reproduces Kerr--Sen from a Kerr seed and
the Ricci flatness of the KLC seed as controls; and, after exact calibration
against the Kretschmann scalars of Schwarzschild, Reissner--Nordstr\"om,
Kerr in two charts, and Kerr--Newman, agrees exactly with the symbolic
curvature invariants at all thirty (KNLC) and twenty-one (KSLC, equatorial)
tested rational points, the smooth-axis limits agreeing to between
$10^{-24}$ and $10^{-42}$.
(iii)~\emph{Symbolic identities.}  The factorization
$\mathcal W=\Sigma\mathcal H$, the $\Sigma$-divisibility of the azimuthal
gauge numerators, the ring, axis, conicity, and flux values, the relations
$\mathsf I^{2}=\mathrm{id}$ and
$\mathsf I\circ\mathsf H_c=\mathsf D_c\circ\mathsf I$, the static-sector
field equations for general $m$ and $q$, the slice factorizations
\eqref{eq:K-KSLC-slices}, the ring value \eqref{eq:K-ring-KNLC}, and the
asymptotic coefficients in
\eqref{eq:K-universal}--\eqref{eq:K-gap} are established symbolically for
general parameters. 
\section{Explicit polynomials}
\label{app:polys}
The KNLC rotation polynomial is
\begin{footnotesize}
\begin{align}
P_{\omega}={}&x^{4}\big[4a^{4}m^{2}-6a^{4}mr+a^{4}q^{2}-8a^{2}m^{3}r
+4a^{2}m^{2}q^{2}+16a^{2}m^{2}r^{2}-8a^{2}mq^{2}r-4a^{2}mr^{3}
\nonumber\\
&\qquad+2a^{2}q^{4}-2a^{2}q^{2}r^{2}-4m^{2}r^{4}-2mq^{4}r
+8mq^{2}r^{3}+2mr^{5}+q^{6}-2q^{4}r^{2}-3q^{2}r^{4}\big]
\nonumber\\
&+x^{2}\big[12a^{4}m^{2}-12a^{4}mr+a^{4}q^{2}-24a^{2}m^{3}r
+12a^{2}m^{2}q^{2}+36a^{2}m^{2}r^{2}-14a^{2}mq^{2}r-24a^{2}mr^{3}
\nonumber\\
&\qquad+a^{2}q^{4}+7a^{2}q^{2}r^{2}+24m^{2}r^{4}-24mq^{2}r^{3}
-12mr^{5}+6q^{4}r^{2}+6q^{2}r^{4}\big]
\nonumber\\
&+4a^{4}m^{2}+2a^{4}mr+a^{4}q^{2}-12a^{2}m^{2}r^{2}
\nonumber\\
&\qquad+6a^{2}mq^{2}r-4a^{2}mr^{3}+a^{2}q^{4}
\nonumber\\
&\qquad+a^{2}q^{2}r^{2}-6mr^{5}.
\label{eq:Pomega}
\end{align}
\end{footnotesize}
The KLC seed-frame polynomial is
\begin{footnotesize}
\begin{align}
G={}&(a^{4}x^{4}-6a^{4}x^{2}-3a^{4})r^{5}
-2(a^{4}mx^{4}-6a^{4}mx^{2}+3a^{4}m+16a^{2}m^{3}-16m^{5})r^{4}
\nonumber\\
&-2a^{6}(x^{4}+6x^{2}+1)r^{3}
+4(2a^{6}mx^{4}+3a^{6}mx^{2}-3a^{6}m-8a^{4}m^{3}x^{2}-8a^{4}m^{3}
\nonumber\\
&\qquad+8a^{2}m^{5}x^{2}+8a^{2}m^{5})r^{2}
+(-3a^{8}x^{4}-6a^{8}x^{2}+a^{8}-4a^{6}m^{2}x^{4}-12a^{6}m^{2}
\nonumber\\
&\qquad+64a^{4}m^{4}x^{2}-64a^{4}m^{4}-64a^{2}m^{6}x^{2}
+64a^{2}m^{6})r
+2a^{8}m(x^{4}+1)-32a^{6}m^{3}x^{2}+32a^{4}m^{5}x^{2}.
\label{eq:Gpoly}
\end{align}
\end{footnotesize}
At $q=0$ the two rotation frames obey
\begin{equation}
 -\frac{aP_\omega}{N}=-\frac{2mG}{a^{3}N}
 -\frac{4m^{2}(3a^{4}+16a^{2}m^{2}-16m^{4})}{a^{3}}.
 \label{eq:frame-relation}
\end{equation}
The polynomial in \eqref{eq:WHfactor}, collected in even powers of $x$, is
\begin{footnotesize}
\begin{align}
\mathcal H={}&4a^{4}m^{2}+4a^{4}mr+a^{4}r^{2}+4a^{2}mr^{3}
+2a^{2}r^{4}+r^{6}
\nonumber\\
&+x^{2}\big[a^{6}+8a^{4}m^{2}-12a^{4}mr+4a^{4}q^{2}
+36a^{2}m^{2}r^{2}-20a^{2}mq^{2}r-12a^{2}mr^{3}
\nonumber\\
&\qquad+4a^{2}q^{4}+6a^{2}q^{2}r^{2}-3a^{2}r^{4}
+2q^{2}r^{4}-2r^{6}\big]
\nonumber\\
&+x^{4}\big[-2a^{6}+4a^{4}m^{2}+12a^{4}mr-6a^{4}q^{2}
-3a^{4}r^{2}-24a^{2}m^{2}r^{2}+24a^{2}mq^{2}r+12a^{2}mr^{3}
\nonumber\\
&\qquad-4a^{2}q^{4}-8a^{2}q^{2}r^{2}+q^{4}r^{2}
-2q^{2}r^{4}+r^{6}\big]
\nonumber\\
&+x^{6}\big[a^{6}-4a^{4}mr+2a^{4}q^{2}+2a^{4}r^{2}
\nonumber\\
&\qquad+4a^{2}m^{2}r^{2}-4a^{2}mq^{2}r-4a^{2}mr^{3}
\nonumber\\
&\qquad+a^{2}q^{4}+2a^{2}q^{2}r^{2}+a^{2}r^{4}\big].
\label{eq:Hpoly}
\end{align}
\end{footnotesize}
The cancellations in the azimuthal potentials are
\begin{equation}
 q(xQ\mathcal N-ar\Dx\mathcal U)=\Sigma\mathcal K_A,
 \qquad
 -q(xQ\mathcal U+ar\Dx\mathcal N)=\Sigma\mathcal K_B,
 \label{eq:KAKBdef}
\end{equation}
where
\begin{align}
\mathcal K_A={}&-aq\big[2a^{2}mx^{2}+2a^{2}m+a^{2}rx^{4}-2a^{2}rx^{2}
+a^{2}r-2mr^{2}x^{4}+6mr^{2}x^{2}
\nonumber\\
&\hspace{35mm}+q^{2}rx^{4}-q^{2}rx^{2}+r^{3}x^{4}-2r^{3}x^{2}+r^{3}\big],
\label{eq:KApoly}\\
\mathcal K_B={}&-qx\big[-a^{4}x^{2}+a^{4}+4a^{2}mrx^{2}-4a^{2}mr
-a^{2}q^{2}x^{2}+2a^{2}q^{2}-2a^{2}r^{2}x^{2}+2a^{2}r^{2}
\nonumber\\
&\hspace{35mm}+q^{2}r^{2}x^{2}-r^{4}x^{2}+r^{4}\big].
\label{eq:KBpoly}
\end{align}
Thus $A_\phi=\mathcal K_A/\mathcal H$ and
$\widetilde A_\phi=\mathcal K_B/\mathcal H$ after cancellation.
\section*{Acknowledgements}
This work was
supported by the Lembaga Penelitian dan Pengabdian kepada Masyarakat,
Universitas Katolik Parahyangan (LPPM-UNPAR).

\section*{Declarations}

\paragraph{Use of generative artificial intelligence.}
During preparation of this manuscript, the author used Claude (Anthropic)
and ChatGPT (OpenAI) for language editing and for checking of symbolic and
numerical code.  The author reviewed and validated all outputs and accepts
full responsibility for the content of the work.

\end{document}